\documentclass[journal,comsoc]{IEEEtran}
\usepackage[T1]{fontenc}
\usepackage{times}
\usepackage{soul}
\usepackage{url}
\usepackage[hidelinks]{hyperref}
\usepackage[utf8]{inputenc}
\usepackage[small]{caption}
\usepackage{graphicx}
\usepackage{amsmath}
\usepackage{amsthm}
\usepackage{booktabs}
\usepackage{algorithm}
\usepackage{algorithmic}
\usepackage{multirow}
\usepackage{threeparttable}
\usepackage{amsfonts}
\usepackage{float}
\usepackage{url}
\usepackage{bm}
\usepackage{xcolor}
\usepackage{subfigure}
\usepackage{amssymb}

\newcommand{\cmmnt}[1]{}

\ifCLASSINFOpdf
\else
\fi
\usepackage{amsmath}
\usepackage[cmintegrals]{newtxmath}
\begin{document}
%
\title{U-Style: Cascading U-nets with Multi-level Speaker and Style Modeling for Zero-Shot Voice Cloning}

%
%


\author{Tao Li,
    Zhichao Wang,
    Xinfa Zhu,
    Jian Cong,
    Qiao Tian, 
    Yuping Wang, 
    Lei Xie~\IEEEmembership{Senior Member,~IEEE}
 \thanks{Corresponding author: Lei Xie}
  
\thanks{Tao Li, Zhichao Wang, Xinfa Zhu, and Lei Xie are with the School of Computer Science, Northwestern Polytechnical University, Xi’an 710072, China. Email: taoli@npu-aslp.org (Tao Li), zcwang\_aslp@mail.nwpu.edu.cn (Zhichao Wang), xfzhu@mail.nwpu.edu.cn (Xinfa Zhu), lxie@nwpu.edu.cn (Lei Xie)}

\thanks{Jian Cong, Qiao Tian, and Yuping Wang are with the Speech, Audio, and Music Intelligence (SAMI) Group, ByteDance, Shanghai 200233, China. Email: congjian.tts@bytedance.com (Jian Cong), tianqiao.wave@bytedance.com (Qiao Tian), wangyuping@bytedance.com (Yuping Wang)}
    }

%

%

\markboth{Journal of \LaTeX\ Class Files,~Vol.~14, No.~8, August~2015}%
{Wang \MakeLowercase{\textit{et al.}}: Bare Demo of IEEEtran.cls for IEEE Communications Society Journals}
%



\maketitle


\begin{abstract}

Zero-shot speaker cloning aims to synthesize speech for any target speaker unseen during TTS system building, given only a single speech reference of the speaker at hand. Although more practical in real applications, the current zero-shot methods still produce speech with undesirable naturalness and speaker similarity. Moreover, endowing the target speaker with arbitrary speaking styles in the zero-shot setup has not been considered. This is because the unique challenge of \textit{zero-shot speaker and style cloning} is to learn the disentangled speaker and style representations from only short references representing an arbitrary speaker and an arbitrary style. 
To address this challenge, we propose \textit{U-Style}, which employs Grad-TTS as the backbone, particularly cascading a \textit{speaker-specific encoder} and a \textit{style-specific encoder} between the text encoder and the diffusion decoder. 
Thus, leveraging signal perturbation, U-Style is explicitly decomposed into speaker- and style-specific modeling parts, achieving better speaker and style disentanglement. 
To improve unseen speaker and style modeling ability, these two encoders conduct multi-level speaker and style modeling by skip-connected U-nets, incorporating the representation extraction and information reconstruction process. 
Besides, to improve the naturalness of synthetic speech, we adopt mean-based instance normalization and style adaptive layer normalization in these encoders to perform representation extraction and condition adaptation, respectively. Experiments show that U-Style significantly surpasses the state-of-the-art methods in unseen speaker cloning regarding naturalness and speaker similarity.
Notably, U-Style can transfer the style from an unseen source speaker to another unseen target speaker, achieving flexible combinations of desired speaker timbre and style in zero-shot voice cloning.

\end{abstract}


%
\IEEEpeerreviewmaketitle

\section{Introduction}
\label{sc:Introduction}

\IEEEPARstart{T}{ext}-to-speech (TTS) systems have made a remarkable process in generating high-quality human-like speech for both the single- and multi-speaker scenarios~\cite{wang2017tacotron,ping2017deep3}.
However, building with the tremendous cost of data collection and annotation, such systems can only generate the voice of speakers \textit{seen} in the training data~\cite{zhou22d_interspeech}, limiting the real-life applications of TTS~\cite{zhou22d_interspeech}.
To mitigate this limitation, \textit{zero-shot speaker cloning}, which aims to generate speech for any speaker \textit{unseen} during TTS model training, based on a single acoustic reference of the speaker, is gaining more and more attention~\cite{xie2021multic}. 

In zero-shot speaker cloning, speaker modeling performance primarily determines the naturalness and speaker similarity of synthetic speech~\cite{choi2022snac}.
Currently, \textit{global} and \textit{multi-level} speaker modeling~\cite{li2022unet} are often used.
For global speaker modeling, the speaker timbre is represented by a fixed-dimensional embedding extracted from a pre-trained speaker verification (SV) model~\cite{arik2018neural,cooper2020zero,zhang2021one} or a jointly trained speaker encoder~\cite{kang2023grad,yourtts,min2021meta}.  
Such global modeling strategies are plagued by poor speaker similarity and usually fail to preserve the \textit{speaker-specific speaking characteristics}, as the voice characteristics of a speaker include not only global timbre information but also local subtle pronunciation variations~\cite{zhou22d_interspeech}. Thus, some studies~\cite{huanggenerspeech,li2022unet,li2022hierarchical} focus on multi-level speaker modeling to capture speaking characteristics at different granularities, which improves speaker similarity by carrying more speaking styles and pronunciation habits.
In parallel, \textit{normalization-based}~\cite{chen2021again,min2021meta} feature extraction and fusion techniques are widely used in both global and multi-level speaker modeling strategies.
However, the naturalness and speaker similarity achieved by existing methods is still undesirable~\cite{wang2023neural}.
Besides, the current zero-shot TTS methods mainly focus on modeling the speaker timbre, and target speakers cannot be assigned a new speaking style. 

\begin{figure}[t]
  \centering\includegraphics[width=\linewidth]{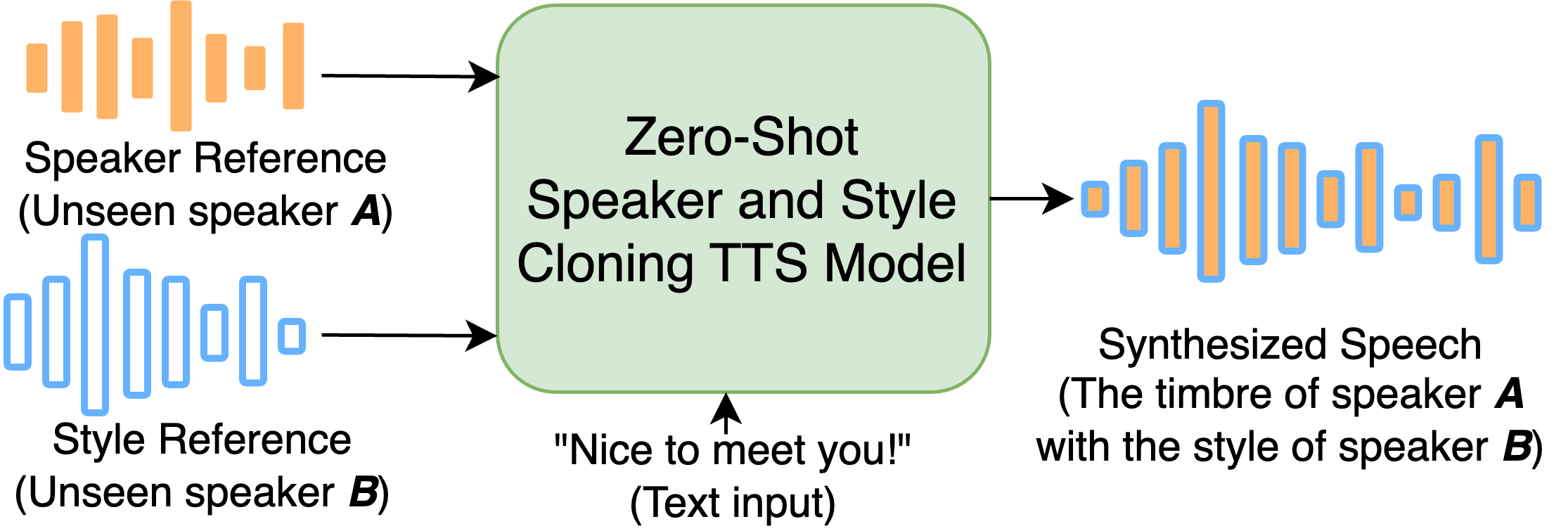}
  \caption{The concept of zero-shot speaker and style cloning, with the aim to build a TTS model capable of synthesizing stylistic speech for unseen speakers, where the style is transferred from another unseen speaker. Here, unseen means no-show in the training set.}
  \label{fig:task}
\end{figure}

Many efforts have been made to synthesize stylistic speech for a target speaker with style transferred from reference speech recorded by another (source) speaker, i.e., the so-called cross-speaker style transfer~\cite{2019Multi}.
The key to achieving cross-speaker style transfer is to extract the \textit{disentangled} speaker and style representations from the reference spectrogram. 
Otherwise, the source speaker information retained in the style representation could reduce the target speaker similarity~\cite {Karlapati2020CopyCatMF}. 
However, the existing cross-speaker style transfer literature~\cite{qiang2023improving} achieves speaker and style disentanglement relying on supervised or semi-supervised strategies with style- and speaker-labeled datasets, limited to transferring pre-defined styles to the target speakers~\cite{kang2023zet}.

In this study, we explore a more challenging scenario, which endows unseen target speakers with arbitrary styles from arbitrary speech references -- \textit{zero-shot speaker and style cloning}, as depicted in its concept in Figure~\ref{fig:task}.
A unique challenge hampers this task: enabling a TTS model to learn disentangled speaker and style representations from the references with both unseen speaker and unseen style.

Assuming that besides linguistic information, \textit{para-linguistic} information in speech can be organized as the timbre of a specific speaker and prosody of a specific style~\cite{valle2020mellotron}, an intuitive way to achieve speaker and style cloning in the zero-shot scenario is to model speaker and style separately.
Following this idea, we propose \textit{U-Style}, taking Grad-TTS~\cite{popov2021grad} as a backbone, inserting a \textit{speaker-specific encoder} and a \textit{style-specific encoder} in a \textit{cascade} manner between the text encoder and diffusion decoder, guiding U-Style to model the speaker- and style-specific information individually and progressively by adopting style-removed timbre-preserved Mel-spectrogram bridging the two encoders.
Here, the style-removed timbre-preserved Mel-spectrogram is achieved by eliminating style information from the original Mel-spectrogram via signal perturbation~\cite{lei2022cross}. 
Importantly, the two encoders utilize a skip-connected U-net to model speaker and style in a \textit{multi-level} manner, taking advantage of layered representation generation and information reconstruction~\cite{li2022unet}.
Specifically, the speaker-specific encoder contains an \textit{extractor} and a \textit{reconstructor}, and the extractor converts the original spectrogram into content representation via block-by-block squeeze-out para-linguistic information, which is used as a multi-level speaker representation that contains more local pronunciation variations.
The reconstructor has an opposite data flow with the extractor to form a skip-connected U-net structure, reusing the multi-level speaker representations at the corresponding blocks to transform the context representation to a style-removed timbre-preserved spectrogram.
Thereby, the speaker-specific encoder is constrained to modeling the speaker timbre only.
The style-specific encoder has the same structural design as the speaker-specific encoder, and it predicts the original spectrogram based on the style-removed timbre-preserved spectrogram, thus formulating an informative prior for the diffusion decoder~\cite{kang2023grad,lee2021priorgrad}.
As the speaker and linguistic information are explicitly fed into the style-specific encoder, it can be guided to model the style-specific information.

We update the normalization strategies in both encoders to improve the speech naturalness and speaker similarity further.
First, all normalization operations only calculate the mean, not the variance, as the mean can reflect global para-linguistic information variations. In contrast, the variance is mainly related to the linguistic content~\cite{li2022hierarchical}, and incorporating it into the multi-level representation may reduce the naturalness of the synthesized speech.
Second, instance normalization (IN) and style adaptive layer normalization (SALN)~\cite{min2021meta} are adopted for the extractor and the reconstructor, respectively, rather than the conventional IN and adaptive instance normalization 
(AdaIN) pair~\cite{chou2019one}.
Compared to layer normalization in SALN, normalization is performed for each instance in AdaIN, which may lose more critical information added previously, compromising the model generalization ability.

Experiments demonstrate that U-Style achieves new state-of-the-art (SOTA) zero-shot speaker cloning performance, and notably, it can conduct zero-shot style cloning to endow unseen target speakers with arbitrary unseen styles.

\begin{figure*}[!htb]
  \centering
  \includegraphics[width=1\linewidth]{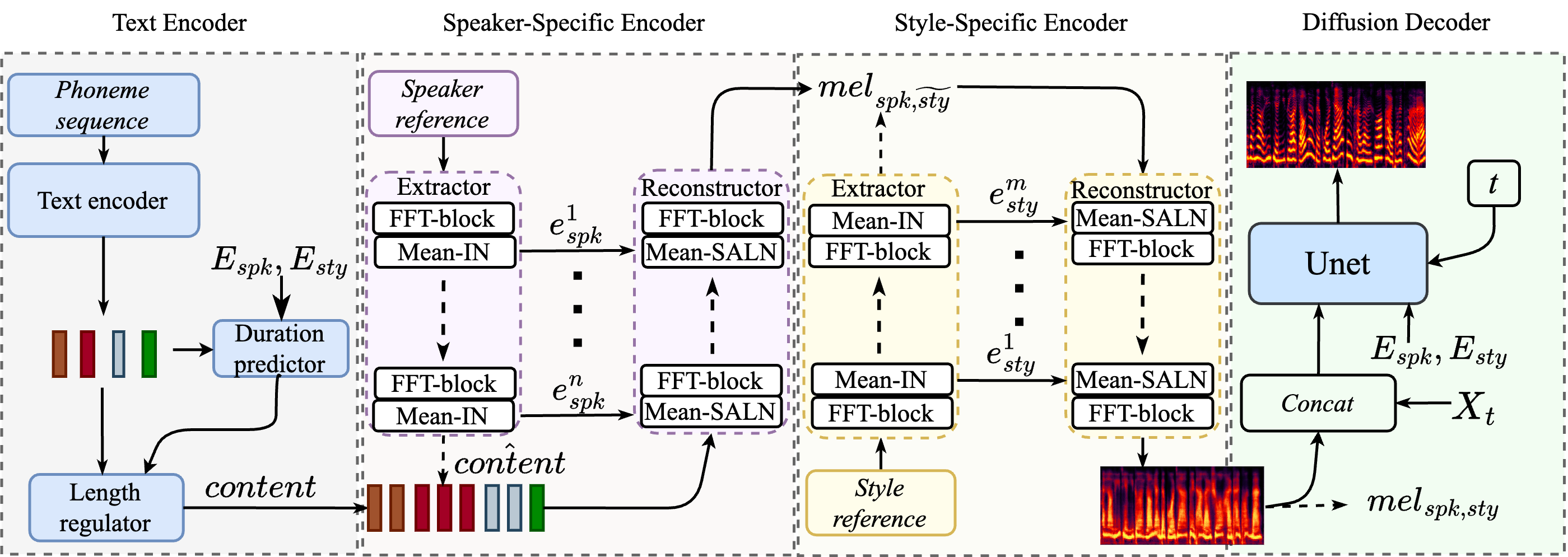}
  \caption{Architecture overview of proposed U-Style.}
  \label{fig:frame1}
\end{figure*}

The rest of this paper is organized as follows. Section~\ref{sc:related work} reviews related work.  Section~\ref{sc:method} introduces the proposed method in detail. Section \ref{sc:experiments} describes the experimental setup.
Section \ref{sc:results} and Section \ref{sc:results2} present the results on zero-shot speaker cloning and zero-shot speaker and style cloning, respectively.  
Finally, the paper concludes in Section \ref{sc:conl}. 
Examples of synthesized speech can be found on the project page\footnote{The synthesized samples can be found at 
https://silyfox.github.io/ustyle/
\label{ft:homepage}}.

\section{Related work}
~\label{sc:related work}
This section briefly introduces the related studies on zero-shot speaker cloning and cross-speaker style transfer. 

\subsection{Zero-shot Speaker Cloning}

Zero-shot speaker cloning aims to generate speech for unseen speakers given only a few seconds of acoustic reference of the speaker~\cite{choi2022snac}.
In recent years, speaker modeling-based approaches have been a common strategy to achieve zero-shot speaker cloning, including \textit{global} and \textit{multi-level} speaker modeling.
Global speaker modeling takes a fixed-dimensional vector as speaker embedding for the TTS model, where the vector is generated by a pre-trained speaker verification (SV)~\cite{Jia2018TransferLF,cooper2020zero,zhang2021one} model or a joint-trained speaker encoder~\cite{kang2023grad,lu2019one,yourtts,min2021meta,casanova2021sc}.
However, due to a single vector's limited information-capturing capability, such global speaker modeling approaches tend to achieve less resemblance between synthetic speech and the speaker's original voice~\cite{zhou22d_interspeech}.

As the voice characteristics of a person include not only global timbre information but also local prosodic variations, some researchers endeavor to capture multi-level speaker representation to improve speaker similarity.
Some works~\cite{fu2019phoneme,zhou22d_interspeech} employed attention mechanisms to capture the content relevance between the reference speech and the input text, generating the fine-grained speaker embedding to transfer the speaker pronunciation characteristics related to phoneme content.
In parallel, a type of progressive multi-level speaker representation~\cite{li2022unet,li2022hierarchical} method is introduced, where an up-sampling stream and a reversely down-sampling stream are constructed to simulate a progressively speaker-adaptation process.
Specifically, the down-sampling stream progressively transforms audio samples into linguistic representation by incorporating the instance normalization (IN)~\cite{ulyanov2016instance} in each block for sample-wise normalization and generates temporal statistics, which is used as multi-level speaker representation. 
Then, the speaker representation is gradually fused into the up-sampling blocks by AdaIN~\cite{chen2021again} to transform linguistic representation back into audio samples.
Recently, VALL-E~\cite{wang2023neural} introduced a neural codec language model (LM) for TTS and employed residual vector quantization (RVQ) to extract multi-level speaker representations.
 NaturalSpeech 2~\cite{shen2023naturalspeech} introduced a speech prompting mechanism to encourage the predicted duration and pitch to follow the diverse speaker information in the speech prompt.
Mega-TTS 2~\cite{jiang2023mega} leveraged a multi-reference timbre encoder to extract timbre information from multiple reference speech.

Although the above literature effectively improved the speaker similarity, how to endow the unseen target speaker with an arbitrary style is not considered.

\subsection{Cross-speaker Style Transfer}
To enable a target speaker to express various styles not present in its training data, existing TTS methods have attempted cross-speaker style transfer, where the style is transferred from another source speaker with the desired style.
The main challenge in this scenario is to disentangle the source speaker's timbre and style to avoid the \textit{speaker leakage} problem~\cite{Karlapati2020CopyCatMF}, which makes synthesized speech sound somehow like it was uttered by the source speaker rather than the target speaker.
Reference-based methods are the mainstream strategy for cross-speaker style transfer, extracting a fixed-length style embedding~\cite{Skerry2018Towards,Wang2018Style,Zhang2019LearningLR,Li2021ControllableET} from a reference speech as the condition for the TTS model. 
However, speaker disentanglement is not well considered in the above methods; speaker and prosodic information are aggregated in the style embedding, resulting in speaker leakage.

To achieve speaker disentanglement, a multi-reference encoder~\cite{Bian2019Multi, 2019Multi} framework can be employed, but the trade-off between speaker timbre and style expressiveness is still inevitable~\cite{li22h_interspeech}. 
In parallel, Li et al.~\cite{Li2021ControllableCE,diclettts} introduces an adversarial-based style disentangling module, constraining the style embedding to be speaker-irrelevant via an orthogonal loss and a gradient reversal layer~\cite{ganin2015unsupervised}.
Zhu et al.~\cite{zhu2023multi} introduces a two-stage framework, adopting multi-label binary vector (MBV) and mutual information (MI) minimization strategies to obtain disentangled representation.
In general, since the disentanglement methods of the above literature rely on supervised or semi-supervised strategies based on labeled data, they can only transfer the styles seen during training to the target speaker but cannot transfer unseen styles.

Compared to the previous works, our study follows the line of reference-based style transfer but attempts to disentangle speaker and style in the zero-shot scenario for the first time, transferring style from an unseen source speaker to another unseen target speaker.

\section{Proposed Approach}\label{sc:method}
\subsection{Overview}
The model structure of our proposed U-Style is illustrated in Figure~\ref{fig:frame1}. 
It adopts Grad-TTS~\cite{popov2021grad} as the backbone while introducing a speaker-specific encoder and a style-specific encoder with acoustic references between the text encoder and diffusion decoder to achieve zero-shot speaker and style cloning.  
We use signal perturbation to obtain style-disentangled representation to guide U-Style to conduct speaker and style disentanglement in the zero-shot scenario.
Specifically, the speaker-specific encoder extracts multi-level speaker representations from the speaker reference to transform the aligned content sequence into a style-removed timbre-preserved spectrogram. 
Correspondingly, the style-specific encoder transforms the style-removed timbre-preserved spectrogram into the original spectrogram based on multi-level style representations extracted from the style reference. 
Besides, we adopt the mean-based instance normalization (IN) and style adaptive layer normalization (SALN) pair in these encoders to improve the naturalness of synthesized speech. 
Finally, guided by the speaker and style information, a diffusion decoder is adopted to restore the target Mel-spectrogram from the terminal distribution formulated by the style-specific encoder.

\subsection{Style-disentangled Representation}
Unlike the previous studies to learn a disentangled latent space for speakers and styles based on corresponding labels in a supervised/semi-supervised manner, for zero-shot scenarios, we employ signal perturbation~\cite{choi2021neural} to achieve speaker and style disentanglement. Specifically, para-linguistic information in human speech, such as speaker timbre and speaking style, is highly dynamic and deeply entangled.
The style is mainly reflected in pitch skeletons and speaking speed, especially in Mandarin, a typical tonal language, while the speaker timbre does not vary with style.
Accordingly, we perturb the pitch information through randomization and monotonization to achieve the style-removed timbre-preserved spectrogram, which serves as a style-disentangled representation to decompose the U-Style into the speaker- and style-specific modeling parts for autonomously learning the disentangled representation.

Formally, given the original spectrogram $mel_{spk, sty}$, where the $spk$ and $sty$ are corresponding speaker timbre and style, respectively. 
The pitch amplitudes are changed by randomly shifting the pitch values, and then all pitch values are monotonized to the median pitch of the utterance to prevent pitch variations within an utterance remaining style-discriminative after pitch shifting.
After the above operations, we can obtain the styleless waveform $mel_{spk, \widetilde{sty}}$, where the speaker timbre $spk$ is preserved, and the process can be expressed as:

\begin{equation}
    mel_{spk, \widetilde{sty}} = \text{F\_{pm}}(\text{F\_{pr}}(mel_{spk, sty})),
\end{equation}
where $\text{F\_{pm}}$ and $\text{F\_{pr}}$ are the pitch randomization and pitch monotonization functions, respectively.

\subsection{Speaker- and Style-specific Modeling}
To make U-Style autonomy learning disentangled speaker and style representations, we take the above style-removed spectrogram $mel_{spk, \widetilde{sty}}$ as an intermediate representation to decompose the model into the speaker-specific and style-specific modeling parts.
As shown in Figure~\ref{fig:frame1}, we introduce a speaker-specific and a style-specific encoder between the text encoder and diffusion decoder. 
During training, the text encoder transforms the phoneme sequence into the aligned content representation $content$ according to its real duration.
The speaker-specific encoder gradually transforms the $content$ to $mel_{spk, \widetilde{sty}}$, which is then transformed to $mel_{spk, sty}$ by the style-specific encoder.
Considering the speaker and style characteristics in human speech naturally include global information and hierarchical variations.
We conduct multi-level modeling with normalization operating in speaker-specific and style-specific encoders to improve generality for unseen speakers and styles.

\subsubsection{Speaker-specific Encoder} 

The speaker-specific encoder consists of an extractor and a reconstructor. They have the same structure and form a skip-connected U-net structure by employing opposite information flows.
The extractor transforms the original spectrogram $mel_{spk, sty}$ to the $content$ by block-by-block with normalization operation to squeeze out the content-irrelevant information, which is reused by the reconstructor with denormalization operation to transform the $content$ to the $mel_{spk, \widetilde{sty}}$.
Such information squeezed out by the extractor can be treated as multi-level speaker representations as the reconstructor is constrained to predict style-removed timbre-preserved $mel_{spk, \widetilde{sty}}$.
Besides, as the target of extractor prediction is aligned phoneme-level representation $content$, the multi-level speaker representation could contain more local pronunciation variations that further improve speaker similarity.
The process of the speaker-specific encoder and corresponding loss function is defined as follows:
\begin{equation}
    content = \text{Text Encoder}(phoneme, E_{spk},E_{sty}), 
\end{equation}

\begin{equation}
    \{\hat{content}, e_{spk}^{i}\} = \text{ Extractor\_{spk}}(mel_{spk,sty}),
\end{equation}

\begin{equation}
   \hat{mel}_{spk, \widetilde{sty}} = \text{Reconstructor\_{spk}}(content, e_{spk}^{i}),
\end{equation}

\begin{equation}
\begin{aligned}
{{\cal L}_{spk}} &= ||\hat{content}-content||_{2} +
\\& ||\hat{mel}_{spk, \widetilde{sty}}- mel_{spk, \widetilde{sty}}||_{2},
\end{aligned}
\end{equation}
where $content$ and $\hat{content}$ are aligned content representations predicted by the text encoder and extractor $\text{Extractor\_{spk}}$. 
$\hat{mel}_{spk, \widetilde{sty}}$ is the speaker-preserved style-removed Mel-spectrogram predicted by the reconstructor $\text{Reconstructor\_{spk}}$.
$E_{spk}$ and $E_{sty}$ are the sum of multi-level speaker representations $e_{spk}^{i}$ and multi-level style representations $e_{sty}^{j}$ (will be introduced later), respectively.

\subsubsection{Style-specific Encoder}
The style-specific encoder has the same pattern as the speaker-specific encoder and also forms a skip-connected Unet-structure via an extractor and reconstructor.
The difference is that it recovers the removed style information in $mel_{spk, \widetilde{sty}}$ according to the extracted multi-level style representation, transforming the  $mel_{spk, \widetilde{sty}}$ predicted by the speaker-specific encoder back to the original spectrogram $mel_{spk, sty}$.
Specifically, the extractor gradually converts the $mel_{spk, sty}$ to the speaker-preserved but style-removed Mel-spectrogram $mel_{spk, \widetilde{sty}}$, so the residual information squeezed out by normalization operation in each block can form a multi-level style representation.
Since the content and speaker information is explicitly provided in $mel_{spk, \widetilde{sty}}$, the reconstructor is constrained to catch style variations.
The process of the style-specific encoder and corresponding loss function is defined as follows:

\begin{equation}
    \{\hat{mel_{spk, \widetilde{sty}}}, e_{sty}^{j}\} = \text{Extractor\_{sty}}(mel_{spk,sty}),
\end{equation}

\begin{equation}
   \hat{mel}_{spk, sty} = \text{Reconstructor\_{sty}}(mel_{spk, \widetilde{sty}}, e_{sty}^{j}),
\end{equation}

\begin{equation}
\begin{aligned}
{{\cal L}_{sty}} = &||\hat{mel}_{spk, \widetilde{sty}}- mel_{spk, \widetilde{sty}}||_{2}
\\&  + ||\hat{mel}_{spk, sty}-mel_{spk, sty}||_{2},
\end{aligned}
\end{equation}
where $\hat{mel}_{spk, \widetilde{sty}}$ and $\hat{mel}_{spk, sty}$ are speaker-preserved style-removed Mel-spectrogram and original spectrogram predicted by the extractor $\text{Extractor\_{sty}}$ and reconstructor $\text{Reconstructor\_{sty}}$.

\subsection{Mean-based Normalization}
As the variance statistics of a speech utterance are mainly related to its linguistic content, adopting the reference's variance as the condition could reduce the naturalness if the reference and input text have inconsistent linguistic content~\cite{2018Hierarchical}. 
For both speaker- and style-specific encoders, we apply mean-based IN behind each feed-forward transformer (FFT) block for the extractor to squeeze out a group instance-specific means as the multi-level representation. 
Unlike the conventional way~\cite{li2022unet,chen2021again}, we utilize mean-based SALN in the reconstructor rather than the AdaIN corresponding to IN operations of the extractor, as IN used in AdaIN constantly removes instance-specific means, which could lose condition information added previously and reduce the adaptation quality. 
Specifically, in the reconstructor, the mean-based SALN is adopted before each FFT block, and the multi-level representations generated by the extractor are reused by these SALNs.

\subsection{Variable Durations}
As the same content deductive with different speakers and styles should have variable durations, the duration predictor in text encoder takes speaker embedding $E_{spk}$ and style embedding $E_{sty}$ as extra input:
\begin{equation}
E_{spk}=\sum_{\substack{i=1}}^n e_{spk}^{i} , \quad
E_{sty}=\sum_{\substack{l=1}}^m e_{sty}^{j} ,
\end{equation}
where $e_{spk}^{i}$ and $e_{sty}^{j}$ are the outputs of the $i$ and $j$ blocks of the extractor in the speaker- and style-specific encoder, respectively.

\subsection{Final Objective Function}
All modules included in the U-Style are joint-trained.
The final objective function to train the proposed U-Style is:
\begin{equation}
    {{\cal L}_{total}} =0.8{{\cal L}_{spk}} + {{\cal L}_{sty}} + {{\cal L}_{diff}} + \mathcal{L}_{dur},
\end{equation}
where $\mathcal{L}_{dur}$ and ${\cal L}_{diff}$ are loss functions of the duration predictor and diffusion decoder, trained by the mean square error (MSE) loss with the ground-truth duration and Mel-spectrogram, respectively.
Specifically, ${\cal L}_{diff}$ of diffusion decoder is defined as follows:

\begin{equation}
\begin{aligned}
\mathcal{L}_{diff}=&\underset{\theta}{\arg \min } \int_{0}^{1} \lambda_{t} \mathbb{E}_{X_{0}, X_{t}}\|s_{\theta}(X_{t}, \hat{mel}_{spk, sty}, t, 
\\ &
E_{spk}, E_{sty}) -\nabla \log p_t(X_{t} \mid X_{0})\|_{2}^{2} d t,
\end{aligned}
\end{equation}
where $\lambda_{t} = 1-e^{-\int_{0}^{t} \beta_{s} ds}, 0<s<t$.
The reverse diffusion parameterized with the diffusion decoder $s_{\theta}$ is trained to approximate gradient of log-density of $X_{t}$ given $X_{0}$, $E_{spk}$, $E_{sty}$ and $\hat{mel}_{spk, sty}$. 
In this case, the diffusion decoder accepts a more informative prior formulated by the style-specific encoder and also takes the $E_{spk}$ and $E_{sty}$ as conditions for each ResBlock in Unet, which can effectively improve the generation quality of the diffusion model~\cite{lee2021priorgrad,diclettts}.

During training, the speaker and style references are ground truth Mel-spectrograms, while during inference, they can be unseen acoustic references with the desired speaker and style, as depicted in Figure~\ref{fig:task}.

\begin{table*}[!htb]
\caption{The speaker similarity, style similarity, and naturalness MOS in zero-shot speaker cloning, with 95\% confidence interval.}
\label{tab:mos}\setlength{\tabcolsep}{1.8mm}
\begin{tabular}{c|ccc|ccc|ccc}
\toprule
\textbf{}   & \multicolumn{3}{c|}{\textbf{Speaker similarity MOS}}  & \multicolumn{3}{c|}{\textbf{Style/Emotion similarity MOS}}  & \multicolumn{3}{c}{\textbf{Naturalness MOS}}  \\ \toprule
\textbf{Testing}  & \textbf{AISHELL-3 } & \textbf{Mspk-CN} & \textbf{ESD/DB\_6}  & \textbf{AISHELL-3 } & \textbf{Mspk-CN} & \textbf{ESD/DB\_6}    & \textbf{AISHELL-3 } & \textbf{Mspk-CN} & \textbf{ESD/DB\_6} \\ \midrule

AnySPK      & 3.78 $\pm$ 0.11 & 3.81 $\pm$ 0.08 & 3.68 $\pm$ 0.10 
            & 3.69 $\pm$ 0.05 & 3.52 $\pm$ 0.08 & 3.58 $\pm$ 0.05 
            & \textbf{4.19 $\pm$ 0.09} & \textbf{4.16 $\pm$ 0.07} & \textbf{4.07 $\pm$ 0.07} \\ 

Unet-TTS    & 4.15 $\pm$ 0.11   & 4.14 $\pm$ 0.15 &4.11 $\pm$ 0.12 
            & 4.09 $\pm$ 0.09 & 4.16 $\pm$ 0.04 & 4.15 $\pm$ 0.06
            & 3.77 $\pm$ 0.07 & 3.74 $\pm$ 0.11 & 3.69 $\pm$ 0.08 \\ 

VALL-E       & \textbf{4.23 $\pm$ 0.08}   & \textbf{4.29 $\pm$ 0.11} & \textbf{4.25 $\pm$ 0.08} 
           & 4.06 $\pm$ 0.08 & 4.08 $\pm$ 0.11 & 4.10 $\pm$ 0.07 
            & 3.89 $\pm$ 0.11 & 3.83 $\pm$ 0.13 & 3.78 $\pm$ 0.07 \\ \toprule

U-Style-100     &4.16 $\pm$ 0.07   & 4.20 $\pm$ 0.13 & 4.19 $\pm$ 0.11 
            & 4.10 $\pm$ 0.10 & 4.13 $\pm$ 0.11 & 4.12 $\pm$ 0.09 
            &  4.13 $\pm$ 0.04 & 4.09 $\pm$ 0.07 & \textbf{4.01 $\pm$ 0.09} \\
            
U-Style-500     &\textbf{4.20 $\pm$ 0.04}   & \textbf{4.26 $\pm$ 0.11} & \textbf{4.23 $\pm$ 0.09} 
            & \textbf{4.16 $\pm$ 0.10} & \textbf{4.23 $\pm$ 0.06} & \textbf{4.20 $\pm$ 0.06} 
            &  \textbf{4.15 $\pm$ 0.05} & \textbf{4.10 $\pm$ 0.11} & 3.98 $\pm$ 0.08 \\
            
\bottomrule
\end{tabular}
\end{table*}

\section{Experiments}
\label{sc:experiments}

\subsection{Model Configurations}
\label{sc:model_config}

The text encoder has the same architecture in DelightfulTTS~\cite{liu2021delightfultts}, which is composed of a pre-net, six conformer blocks~\cite{gulati2020conformer} with multi-head self-attention, and the final linear projection layer to generate 448-dimensional linguistic representation. 
Speaker-specific and style-specific encoders have the same structure but differ in exact parameter settings.
Specifically, the speaker-specific encoder's extractor and the reconstructor are stacked with five FFT blocks, while four FFT blocks are used in the style-specific encoder.
The diffusion decoder's architecture is based on U-net and is the same as in DiCLET-TTS~\cite{diclettts}.
Finally, U-Style has 89M parameters in its model size.

\subsection{Training Dataset}
To verify the influence of data size on the generation performance, we took two different scales of datasets for training:

\begin{itemize}
\item \textbf{100-hour} consists of AISHELL-3~\cite{zhou22d_interspeech} and Mspk-CN datasets, including 243 distinct speakers. 
AISHELL-3 is an open-source multi-speaker speech corpus containing 85 hours of recordings spoken by 218 native Mandarin speakers. 
Mspk-CN is a multi-speaker corpus containing 20 hours of recordings spoken by 25 native Mandarin speakers with multiple speaking styles. 
We randomly split eight speakers (gender balance) from AISHELL-3 and Mspk-CN, respectively, as unseen speakers for evaluation, and the remaining 227 speakers are used for training, containing over 91k utterances.
 
\item \textbf{500-hour} includes the above 100-hour dataset and an extra 400 hours of multi-speaker Mandarin speech, containing over 510k utterances from 478 distinct speakers.
\end{itemize}

\subsection{Evaluation Dataset}

To verify the generation performance of U-Style respectively on \textit{neutral}, \textit{stylistic}, and \textit{emotional} speakers, besides the 16 neutral and stylistic speakers separated from AISHELL-3 and Mspk-CN, we also take the speakers from the emotional speech dataset (ESD)~\cite {zhou2021seen} and DB\_6~\cite {Li2021ControllableCE} as unseen emotional speakers.
Specifically, ten Mandarin speakers (gender-balanced) of five emotion types in ESD were used for evaluation, while DB\_6 is a single-speaker emotion dataset containing six distinct emotions.

\subsection{Training Setups}
Eighty-dimension Mel-spectrogram represents waveform with a frame length of 50ms, frame-shift of 12.5ms, hop size of $200$, and $16$ KHz sampling rate.
A grapheme-to-phoneme (G2P) module converts text sentences into phoneme sequences.
The phoneme duration is obtained by a pre-trained Montreal Forced Alignment (MFA) tool~\cite{Ren2019FastSpeechFR}.
We train the models for 400K iterations with a batch size of 14 on 8 NVIDIA Tesla V100 GPUs.
Finally, a well-trained Hifi-GAN~\cite{kong2020hifi} is adopted as the neural vocoder to the reconstruct waveform from the predicted Mel-spectrogram.

\subsection{Comparison Methods}
We selected the most recent relevant methods to compare with our proposed U-Style: 
\begin{itemize}
\item \textbf{AnySPK}~\cite{kang2023grad}: a Grad-TTS based method utilizes the Mel-style encoder~\cite{min2021meta} and style-adaptive encoder to achieve global speaker modeling.
\item \textbf{Unet-TTS}~\cite{li2022unet}: a multi-level speaker modeling method employs a CNN-based style encoder, which forms a U-net structure with the decoder. 
\item \textbf{VALL-E}~\cite{wang2023neural}: a language-model-based method employs an RVQ module to achieve multi-level speaker modeling. 
As LM-based approaches are data-hungry to obtain stable performance~\cite{zhang2023speak}, we re-implement VALL-E on 5000 hours of Mandarin speech, including the above 500-hour dataset and 4500 hours of extra data collected from the Internet.
\item \textbf{U-Style-100}: the proposed U-Style trained on the 100-hour dataset.
\item \textbf{U-Style-500}: the proposed U-Style trained on the 500-hour dataset.
\end{itemize}
We re-implement AnySPK and Unet-TTS on the 500-hour dataset for a fair comparison.

\subsection{Evaluation Metrics}
We conduct mean opinion score (MOS)~\cite{zhu2023multi} test for all compared models, where the participants are asked to rate given speech a score ranging from $1$ to $5$ based on the specific purpose.
Corresponding rating criteria is: \textit{bad = 1}; \textit{poor = 2}; \textit{fair = 3}; \textit{good = 4}; \textit{great = 5}, in 0.5 point increments.
AB preference test~\cite{an2022disentangling} is also adopted in the evaluation, and participants are asked to assess the similarity between the synthesized sample with the speaker and style references according to a certain requirement.
Specifically, given ten reserved transcripts for each style, we randomly select references with the corresponding style to generate samples respectively for each speaker.
Thirty participants were asked to evaluate the samples on \textit{speaker similarity}, \textit{style similarity}, and \textit{naturalness}.
Besides, we adopt pitch trajectory visualization to evaluate the style similarity objectively.

\section{Zero-shot speaker cloning results}~\label{sc:results}
In this section, we verify the performance of U-Style on \textit{zero-shot speaker cloning}, where the speaker reference and the style reference are the \textbf{same}, i.e., from the same utterance.

\subsection{Speaker Similarity}
As shown in Table~\ref{tab:mos}, AnySPK has an inferior speaker cloning ability than the other methods, especially for the emotional speakers from ESD.  
We observe that AnySPK often failed to preserve speaker-specific characteristics (such as accent), which could be caused by the insufficient ability of single embedding to capture the dynamically varying speaker characteristics.
VALL-E achieves the best speaker similarity in all three datasets, which is understandable, as it benefits from the size of training data, and the timbre space built by VALL-E is more abundant.
We also notice that the gap between U-Style-500 and VALL-E is not significant, and the speaker similarity from the stylistic and emotional reference is slightly higher than that from natural reference, which indicates that the multi-level speaker representation captures more speaking characteristics from the expressiveness voice and helps improve speaker similarity from the sense of listening.
On the other hand, U-Style-100 scores between Unet-TTS and U-Style-500, indicate that U-Style can already achieve good speaker similarity on the 100 hours of training data. 
We believe that this benefit comes not only from mean-based normalization and de-normalization but also from the fact that we stack the U-net via transformer blocks with stronger modeling capabilities.

\subsection{Style Similarity}
It is necessary to evaluate the style similarity further, as it reflects the model's ability to capture the speaking characteristics of unseen speakers.
Table~\ref{tab:mos} shows that the three multi-level speaker modeling methods are far superior to the global modeling method AnySPK. 
On the other hand, Unet-TTS, U-Style-100, and U-Style-500 are superior to VALL-E in style preservation, possibly since in Unet-TTS and U-Style, the process of multi-level representation extraction and the reconstruction is incorporated together, which could improve the model's ability to capture diverse information from references.
Specifically, U-Style-500 achieves the best style similarity, which we believe benefits from using multi-level modeling to model style-specific information individually, and also employs the M-SALN in the reconstruction process, achieving a better prosody cloning performance than AdaIN.
The diffusion decoder also improves the modeling ability of unseen speakers~\cite{kang2023grad}.

\begin{figure*}[ht]
  \centering
  \includegraphics[width=1\linewidth]{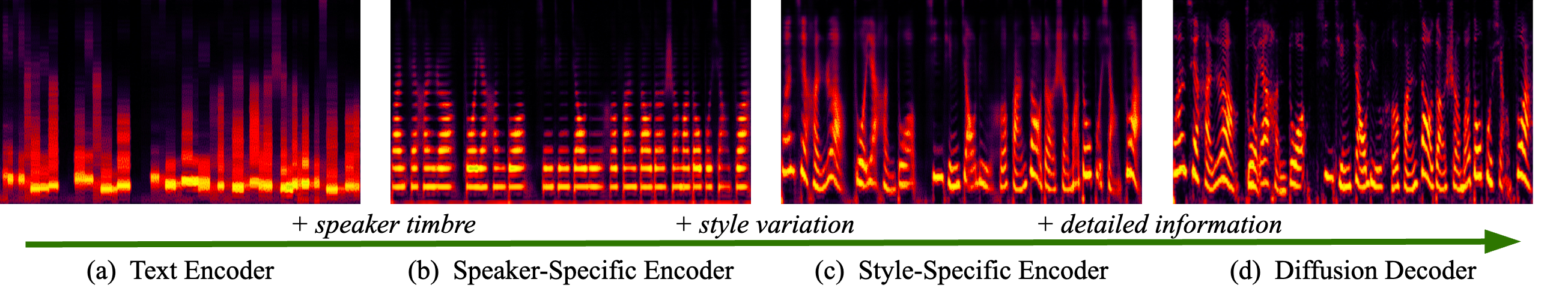}
  \caption{Visualization of the latent representation predicted from each modeling part with a different purpose in U-Style.}
  \label{fig:dataflow}
\end{figure*}

\subsection{Speech Naturalness}
As shown in Table~\ref{tab:mos}, AnySPK achieves good performance in naturalness MOS while considering its poor performance in speaker and style similarity, indicates that the speaker embedding extracted in AnySPK is an average representation of para-linguistic information in the reference, with little impact on the naturalness of synthesized speech.
We observe the same problem for VALL-E in the paper~\cite{wang2023neural}: words are missing or repeated in synthetic speech, especially the reference containing obvious style or accent.
The Unet-TTS achieves the worst naturalness MOS, which seems a trade-off: it could reproduce the reference style but has limited adaptation stability. 
This phenomenon could be caused by the speaker representation in Unet-TTS containing the reference's variance, which is mainly related to content information and could degrade the naturalness in un-parallel inference.
The U-Style-500 achieves slightly decreased MOS compared with AnySPK, but the gap is insignificant.
The above results indicate that U-Style can effectively capture unseen speakers' timbre and style characteristics, allowing the adoption of expressive reference to guide the robust zero-shot speaker cloning of arbitrary text.

Based on the results of speaker similarity, style similarity, and naturalness MOS in Table~\ref{tab:mos}, it can be seen that for U-Style, while an increased data size benefits establishing a more ample timbre and style space to improve the similarity of speaker timbre and style, we observe a compromise between style reproduction and naturalness during the experimental process.
Specifically, in case the acoustic reference contains exaggerated prosodic variation, more detailed style reproduction is beneficial for improving the speaker similarity but not conducive to the naturalness of the synthesized speech, especially in the un-parallel generation when the intonation of the input text content does not match the prosodic variation of the reference.
For example, compared with U-Style-100, although U-Style-500 has advantages in speaker and style similarity, it does not show a significant advantage in naturalness, and the U-Style-100 even shows slightly better naturalness when the reference comes from the emotional dataset ESD.
These experimental results indicate that excessive training data may lead to performance degradation in certain aspects for a model with regular size parameters such as U-Style. 
According to the model's overall performance in different data sizes, we adopt the U-Style-500 model for subsequent experiments, and the subsequent U-Style-500 is referred to as U-Style for short.

\begin{figure}[ht]
	\centering 
	\centerline{\includegraphics[width=0.4\textwidth]{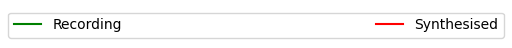}}
	\begin{minipage}{\linewidth}
		\begin{minipage}{0.495\linewidth}
			\includegraphics[width=\textwidth]{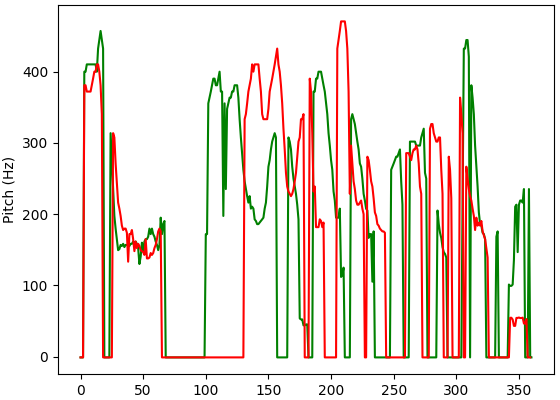}
			\centerline{(a) Angry }
		\end{minipage}
		\hfill
		\begin{minipage}{0.495\linewidth}
			\includegraphics[width=\textwidth]{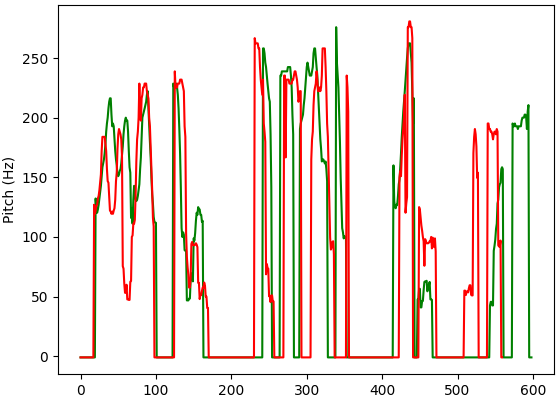}
			\centerline{(b) Tang Poetry}
		\end{minipage}
		\hfill
		\begin{minipage}{0.495\linewidth}
			\includegraphics[width=\textwidth]{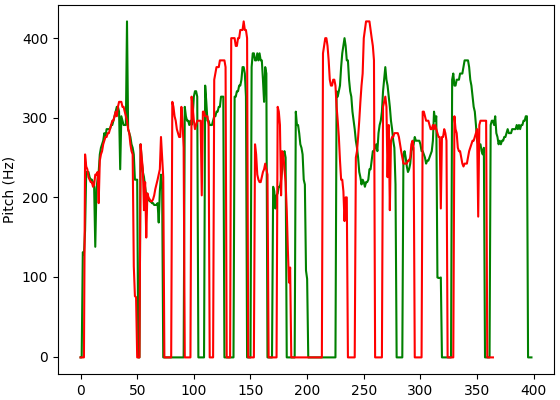}
			\centerline{(c) Happy}
		\end{minipage}
		\hfill
		\begin{minipage}{0.495\linewidth}
			\includegraphics[width=\textwidth]{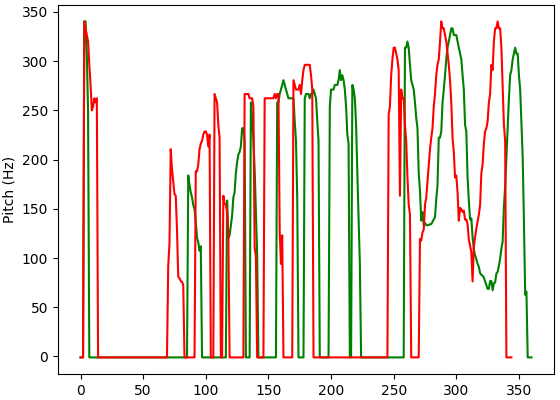}
			\centerline{(d) Martial-arts Novel}
		\end{minipage}
		\hfill
		\begin{minipage}{0.495\linewidth}
			\includegraphics[width=\textwidth]{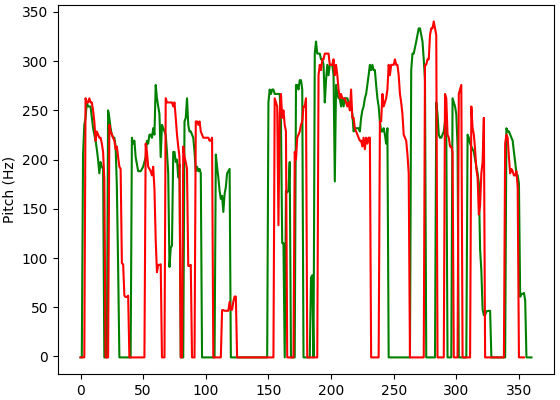}
			\centerline{(e) Fear}
		\end{minipage}
		\hfill
		\begin{minipage}{0.495\linewidth}
			\includegraphics[width=\textwidth]{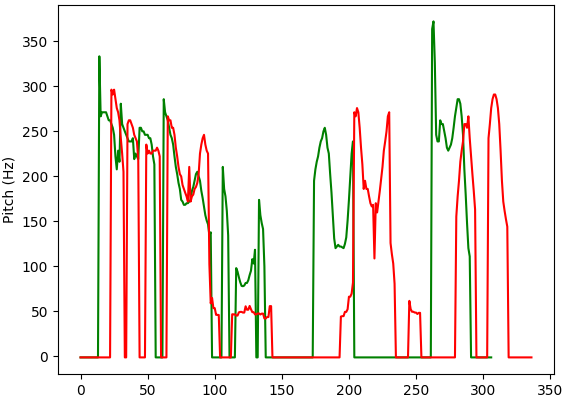}
			\centerline{(f) Humorous}
		\end{minipage}
		\vfill
	\end{minipage}	
	\caption{Pitch visualization results for zero-shot speaker cloning, where each speaker has a different style.}
	\label{fig:pitch_spk}
\end{figure}

\subsection{Visualization of Intermediate Representation}
We visualize the Mel-spectrogram predicted by the different modeling parts with different purposes in U-Style to demonstrate how U-Style gradually transforms the input phoneme sequence into the final Mel-spectrogram.
As shown in Fig~\ref{fig:dataflow}, the subplot (a), (b), (c), and (d) correspond to the output of the text encoder, speaker-specific encoder, style-specific encoder, and diffusion decoder, respectively.
As expected, the text encoder maps the input phoneme sequence to a length-regulated content Mel-spectrogram, which the speaker-specific encoder transforms into a timbre-related but style-removed Mel-spectrogram $mel_{spk, \widetilde{sty}}$ with a smooth and invariable pitch curve.
Subsequently, the style-specific encoder endows prosodic variation to $mel_{spk, \widetilde{sty}}$ to obtain $mel_{spk, sty}$, then the diffusion decoder further refines detailed information. 

The above results indicate that the speaker- and style-specific encoders in U-Style can model the timbre- and style-specific information separately without disturbing the content, and the diffusion decoder helps to refine the detailed information to improve the synthesized speech expressiveness further.

\subsection{Pitch Trajectories Visualization} 
As mentioned earlier, the speaking style characteristics of a person are highly related to pitch trend and duration, especially in Mandarin, a typical tonal language.
Here, in addition to the subjective evaluation, the pitch curves of the speech synthesized by the U-Style via different unseen speakers with different styles are plotted. 
The visualized results are shown in Fig~\ref{fig:pitch_spk}, in which the pitch trajectory of each synthetic speech is drawn based on the Mel-spectrograms.
Specifically, we randomly selected six distinct styles: \textit{Anger}, \textit{Tang Poetry}, \textit{Happy}, \textit{Martial-arts Novel}, \textit{Fear}, and \textit{Humorous}.
For the convenience of comparison, for each style, the content of the input text, speaker reference, and style reference are all consistent.

As we can see, in each subplot, the pitch trajectory of the U-Style synthesized speech closely matches the reference speech pattern.
For detailed characteristics, taking \textit{Fear} as an example, we can find that the synthetic speech and the reference both have slow speech rates and have nearly consistent pitch variety.
For overall characteristics, taking \textit{Martial-arts Novel} as an example, the synthetic speech and the reference have the same overall trend: the pitch is getting higher and higher.
The above subjective and objective results indicate that the U-Style can successfully clone unseen speakers' speaking style and timbre characteristics while obtaining appropriate naturalness.

\section{Zero-shot speaker and style cloning results}~\label{sc:results2}
The previous subsection proved that U-Style could obtain state-of-the-art (SOTA) zero-shot speaker cloning performance.
We proceed to verify the performance of U-Style in zero-shot speaker and style cloning, i.e., endowing target speakers (given speaker reference) with arbitrary styles (given style reference) from another source speaker.

\begin{table}[!t]
\caption{The naturalness MOS in zero-shot speaker and style cloning, with 95\% confidence interval.}
\label{tab:mos1}
\setlength{\tabcolsep}{4mm}
\begin{tabular}{c|c|c}
\toprule
   & \multicolumn{1}{c|}{\textbf{Style reference}}  & \multicolumn{1}{c}{\textbf{Emotion reference}} \\ \toprule
\textbf{Speaker reference}  & Mspk-CN & ESD/DB\_6 \\ \midrule

AISHELL-3        &3.64 $\pm$ 0.09          &3.60 $\pm$ 0.08  \\ 

Mspk-CN    &\textbf{3.72 $\pm$ 0.04}  &3.55 $\pm$ 0.07 \\ 

ESD         & 3.57 $\pm$ 0.10         &\textbf{3.66 $\pm$ 0.05} \\\midrule
Average     &3.64$\pm$0.08  &3.60$\pm$0.06 \\ 
\bottomrule
\end{tabular}
\end{table}

\subsection{Speech Naturalness}
First, we evaluate naturalness, where the style reference is randomly selected from expressiveness datasets, Mspk-CN, ESD, and DB\_6, as the speakers in AISHELL-3  have a neutral style.
As shown in table~\ref{tab:mos1}, the naturalness of the synthesized speech is highest when the speaker and style references come from the same dataset.
This is understandable since the reference speech from the same dataset is acoustically closer.
As we can see, style references from Mspk-CN achieved higher naturalness MOS than ESD ones.
Similarly, we also observe that emotional references from DB\_6 achieved higher naturalness MOS than ESD.
This phenomenon could be because the ESD has poor speech quality~\cite{li2022unet}, including some noise and reverberation.
Overall, for zero-shot speaker and style cloning, the synthesized speech of U-Style achieves acceptable naturalness of the synthesized speech, and the average score is close to \textit{good}.
These results demonstrate that U-Style can generate reasonable and intelligible speech for arbitrary text when speakers and style references have different speakers and styles. 

\begin{figure}[h]
  \centering
  \includegraphics[width=\linewidth]{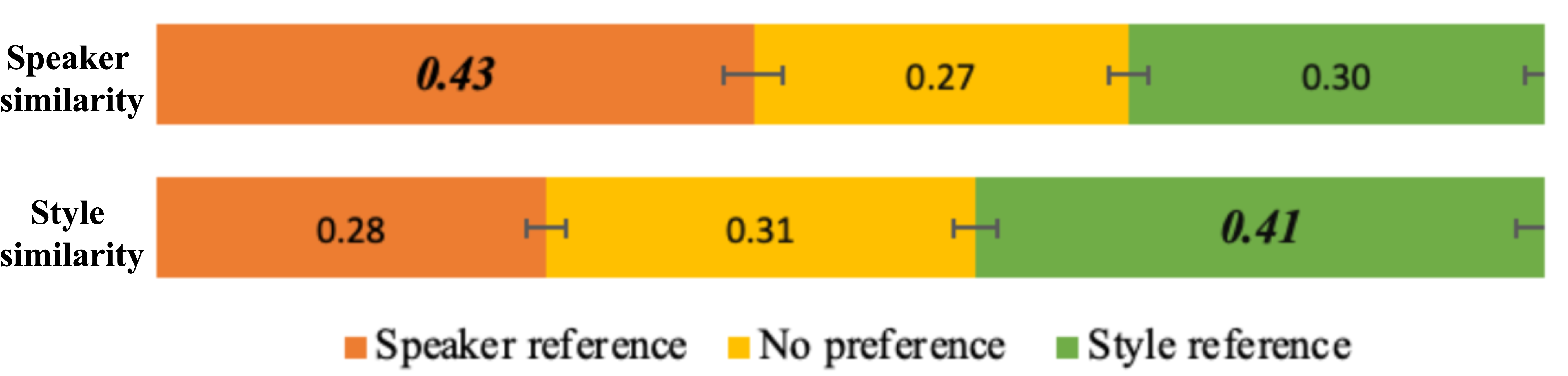}
  \caption{Speaker similarity and style similarity preferences in zero-shot speaker and style cloning.}
  \label{fig:ab}
\end{figure}

\subsection{Speaker and Style Preference}
To investigate U-Style preference on target speaker timbre retention and source style simulation.
We conduct an AB preference test on speaker similarity between synthetic speech and speaker reference and another AB preference test on style similarity between synthetic speech and style reference. The results are shown in Fig.~\ref{fig:ab}.
Regarding speaker and style similarity, as expected, most listeners perceived the synthetic speech to be more similar to the speaker reference in timbre and the style reference in style expressive.
We observe that the synthetic speech can preserve some speaking characteristics of the target speaker, such as accent, which is critical for speaker identification after altering style.
This phenomenon shows that the speaker-specific encoder in U-Style can remove style information well but retain some speaker-specific speaking characteristics. Meanwhile, the style-specific encoder has a limited impact on speaker timbre while modeling style variations.

\subsection{Case Analysis}
\begin{figure}[h]
	\centering 
	\centerline{\includegraphics[width=0.4\textwidth]{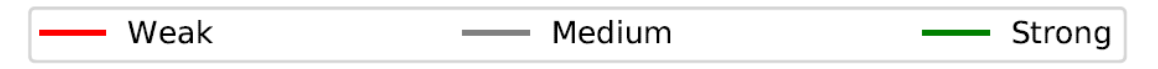}}
	\begin{minipage}{\linewidth}
		\begin{minipage}{0.495\linewidth}
			\includegraphics[width=\textwidth]{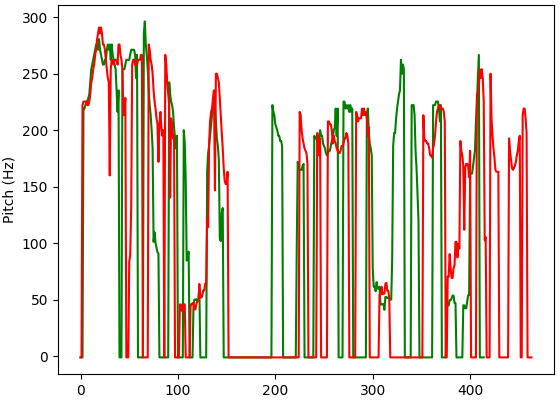}
			\centerline{(a) Sad }
		\end{minipage}
		\hfill
		\begin{minipage}{0.495\linewidth}
			\includegraphics[width=\textwidth]{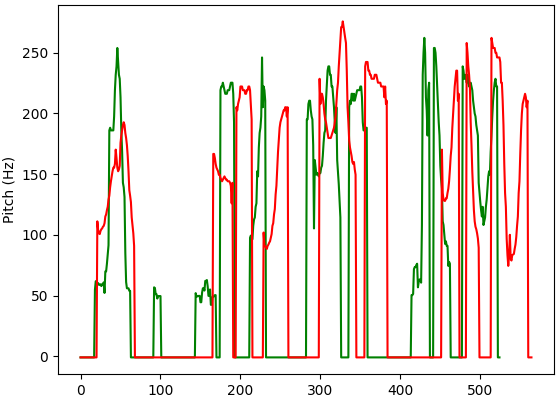}
			\centerline{(b) Tang Poetry}
		\end{minipage}
		\vfill
	\end{minipage}	
	\caption{Pitch visualization for zero-shot speaker and style cloning, where speaker and style are derived from different unseen speakers.}
	\label{fig:pitch_spksty}
\end{figure}

To further demonstrate the influence of style reference on synthesized speech, we also plot two examples of pitch curves for the synthetic \textit{Sad} and \textit{Tang Poetry} samples with random neutral target speakers in Fig.~\ref{fig:pitch_spksty}.
The content of the input text is consistent with the style reference for comparison convenience.
Comparing the ground true style reference and generated speech by U-Style, we can find that although there is a difference in speaking rate, they have a significantly correlated pitch contour.

Specifically, the speaking rate is also affected by speaker reference as the duration predictor also takes the speaker embedding as input, according to the fact that the styles deduced by different speakers should have variable duration.
It indicates that the U-Style can model speaker-specific and style-specific information individually, as expected.

The above results demonstrate that the U-Style can disentangle the speaker and style in the zero-shot scenario, transferring the style from an unseen source speaker to another unseen target speaker.
However, regarding overall performance in speaker and style cloning, U-Style still has room to improve the naturalness of synthesized speech and the trade-off between timbre preservation and expressiveness reproduction.
In particular, we observed during our experiments that the timbre leakage problem~\cite{Karlapati2020CopyCatMF} could be more severe when the gender of the speaker reference and style reference are inconsistent.

\section{CONCLUSIONS}~\label{sc:conl}

This paper proposes U-Style to explore \textit{zero-shot speaker and style cloning}, endowing unseen target speakers with arbitrary styles transferred from another unseen source speaker.
Based on the signal perturbation, U-Style is divided into speaker- and style-specific modeling parts to automatically guide the model to learn disentangled speaker and style representations.
U-Style used a skip-connected U-net structure with M-IN and M-SALN pairing strategies for multi-level speaker and style modeling to achieve better speaker and style similarity.
Besides, the terminal distribution of the forward diffusion is parameterized into a more informative prior that effectively improves the generation quality and efficiency of the diffusion model.
Experimental results show that U-Style achieved SOTA zero-shot speaker cloning performance and produced reasonable performance on both naturalness and similarity in zero-shot speaker and style cloning.   

\bibliographystyle{IEEEtran}
\bibliography{mybibfile.bib}

\begin{thebibliography}{10}
\providecommand{\url}[1]{#1}
\csname url@samestyle\endcsname
\providecommand{\newblock}{\relax}
\providecommand{\bibinfo}[2]{#2}
\providecommand{\BIBentrySTDinterwordspacing}{\spaceskip=0pt\relax}
\providecommand{\BIBentryALTinterwordstretchfactor}{4}
\providecommand{\BIBentryALTinterwordspacing}{\spaceskip=\fontdimen2\font plus
\BIBentryALTinterwordstretchfactor\fontdimen3\font minus
  \fontdimen4\font\relax}
\providecommand{\BIBforeignlanguage}[2]{{%
\expandafter\ifx\csname l@#1\endcsname\relax
\typeout{** WARNING: IEEEtran.bst: No hyphenation pattern has been}%
\typeout{** loaded for the language `#1'. Using the pattern for}%
\typeout{** the default language instead.}%
\else
\language=\csname l@#1\endcsname
\fi
#2}}
\providecommand{\BIBdecl}{\relax}
\BIBdecl

\bibitem{wang2017tacotron}
Y.~Wang, R.~Skerry-Ryan, D.~Stanton, Y.~Wu, R.~J. Weiss, N.~Jaitly, Z.~Yang,
  Y.~Xiao, Z.~Chen, S.~Bengio \emph{et~al.}, ``Tacotron: Towards end-to-end
  speech synthesis,'' \emph{Proc. Interspeech 2017}, pp. 4006--4010, 2017.

\bibitem{ping2017deep3}
W.~Ping, K.~Peng, A.~Gibiansky, S.~{\"O}. Arik, A.~Kannan, S.~Narang,
  J.~Raiman, and J.~Miller, ``Deep voice 3: 2000-speaker neural
  text-to-speech.'' \emph{arXiv preprint arXiv:1710.07654}, 2017.

\bibitem{zhou22d_interspeech}
Y.~Zhou, C.~Song, X.~Li, L.~Zhang, Z.~Wu, Y.~Bian, D.~Su, and H.~Meng,
  ``{Content-Dependent Fine-Grained Speaker Embedding for Zero-Shot Speaker
  Adaptation in Text-to-Speech Synthesis},'' in \emph{Proc. Interspeech 2022},
  2022, pp. 2573--2577.

\bibitem{xie2021multic}
Q.~Xie, X.~Tian, G.~Liu, K.~Song, L.~Xie, Z.~Wu, H.~Li, S.~Shi, H.~Li, F.~Hong
  \emph{et~al.}, ``The multi-speaker multi-style voice cloning challenge
  2021,'' in \emph{ICASSP 2021-2021 IEEE International Conference on Acoustics,
  Speech and Signal Processing (ICASSP)}.\hskip 1em plus 0.5em minus
  0.4em\relax IEEE, 2021, pp. 8613--8617.

\bibitem{choi2022snac}
B.~J. Choi, M.~Jeong, J.~Y. Lee, and N.~S. Kim, ``Snac: Speaker-normalized
  affine coupling layer in flow-based architecture for zero-shot multi-speaker
  text-to-speech,'' \emph{IEEE Signal Processing Letters}, vol.~29, pp.
  2502--2506, 2022.

\bibitem{li2022unet}
R.~Li, D.~Pu, M.~Huang, and B.~Huang, ``Unet-tts: Improving unseen speaker and
  style transfer in one-shot voice cloning,'' in \emph{ICASSP 2022-2022 IEEE
  International Conference on Acoustics, Speech and Signal Processing
  (ICASSP)}.\hskip 1em plus 0.5em minus 0.4em\relax IEEE, 2022, pp. 8327--8331.

\bibitem{arik2018neural}
S.~Arik, J.~Chen, K.~Peng, W.~Ping, and Y.~Zhou, ``Neural voice cloning with a
  few samples,'' \emph{Advances in neural information processing systems},
  vol.~31, 2018.

\bibitem{cooper2020zero}
E.~Cooper, C.-I. Lai, Y.~Yasuda, F.~Fang, X.~Wang, N.~Chen, and J.~Yamagishi,
  ``Zero-shot multi-speaker text-to-speech with state-of-the-art neural speaker
  embeddings,'' in \emph{ICASSP 2020-2020 IEEE International Conference on
  Acoustics, Speech and Signal Processing (ICASSP)}.\hskip 1em plus 0.5em minus
  0.4em\relax IEEE, 2020, pp. 6184--6188.

\bibitem{zhang2021one}
Y.~Zhang, H.~Che, J.~Li, C.~Li, X.~Wang, and Z.~Wang, ``One-shot voice
  conversion based on speaker aware module,'' in \emph{ICASSP 2021-2021 IEEE
  International Conference on Acoustics, Speech and Signal Processing
  (ICASSP)}.\hskip 1em plus 0.5em minus 0.4em\relax IEEE, 2021, pp. 5959--5963.

\bibitem{kang2023grad}
M.~Kang, D.~Min, and S.~J. Hwang, ``Grad-stylespeech: Any-speaker adaptive
  text-to-speech synthesis with diffusion models,'' in \emph{ICASSP 2023-2023
  IEEE International Conference on Acoustics, Speech and Signal Processing
  (ICASSP)}.\hskip 1em plus 0.5em minus 0.4em\relax IEEE, 2023, pp. 1--5.

\bibitem{yourtts}
E.~Casanova, J.~Weber, C.~D. Shulby, A.~C. Junior, E.~G{\"o}lge, and M.~A.
  Ponti, ``Yourtts: Towards zero-shot multi-speaker tts and zero-shot voice
  conversion for everyone,'' in \emph{International Conference on Machine
  Learning}.\hskip 1em plus 0.5em minus 0.4em\relax PMLR, 2022, pp. 2709--2720.

\bibitem{min2021meta}
D.~Min, D.~B. Lee, E.~Yang, and S.~J. Hwang, ``Meta-stylespeech: Multi-speaker
  adaptive text-to-speech generation,'' in \emph{International Conference on
  Machine Learning}.\hskip 1em plus 0.5em minus 0.4em\relax PMLR, 2021, pp.
  7748--7759.

\bibitem{huanggenerspeech}
R.~Huang, Y.~Ren, J.~Liu, C.~Cui, and Z.~Zhao, ``Generspeech: Towards style
  transfer for generalizable out-of-domain text-to-speech,'' \emph{Advances in
  Neural Information Processing Systems}, vol.~35, pp. 10\,970--10\,983, 2022.

\bibitem{li2022hierarchical}
X.~Li, S.~Liu, and Y.~Shan, ``A hierarchical speaker representation framework
  for one-shot singing voice conversion,'' \emph{arXiv preprint
  arXiv:2206.13762}, 2022.

\bibitem{chen2021again}
Y.-H. Chen, D.-Y. Wu, T.-H. Wu, and H.-y. Lee, ``Again-vc: A one-shot voice
  conversion using activation guidance and adaptive instance normalization,''
  in \emph{ICASSP 2021-2021 IEEE International Conference on Acoustics, Speech
  and Signal Processing (ICASSP)}.\hskip 1em plus 0.5em minus 0.4em\relax IEEE,
  2021, pp. 5954--5958.

\bibitem{wang2023neural}
C.~Wang, S.~Chen, Y.~Wu, Z.~Zhang, L.~Zhou, S.~Liu, Z.~Chen, Y.~Liu, H.~Wang,
  J.~Li \emph{et~al.}, ``Neural codec language models are zero-shot text to
  speech synthesizers,'' \emph{arXiv preprint arXiv:2301.02111}, 2023.

\bibitem{2019Multi}
M.~Whitehill, S.~Ma, D.~McDuff, and Y.~Song, ``Multi-reference neural tts
  stylization with adversarial cycle consistency,'' \emph{Proc. Interspeech
  2020}, pp. 4442--4446, 2020.

\bibitem{Karlapati2020CopyCatMF}
S.~Karlapati, A.~Moinet, A.~Joly, V.~Klimkov, D.~S{\'a}ez-Trigueros, and
  T.~Drugman, ``Copycat: Many-to-many fine-grained prosody transfer for neural
  text-to-speech,'' 2020, pp. 4387--4391.

\bibitem{qiang2023improving}
C.~Qiang, P.~Yang, H.~Che, Y.~Zhang, X.~Wang, and Z.~Wang, ``Improving prosody
  for cross-speaker style transfer by semi-supervised style extractor and
  hierarchical modeling in speech synthesis,'' in \emph{ICASSP 2023-2023 IEEE
  International Conference on Acoustics, Speech and Signal Processing
  (ICASSP)}.\hskip 1em plus 0.5em minus 0.4em\relax IEEE, 2023, pp. 1--5.

\bibitem{kang2023zet}
M.~Kang, W.~Han, S.~J. Hwang, and E.~Yang, ``Zet-speech: Zero-shot adaptive
  emotion-controllable text-to-speech synthesis with diffusion and style-based
  models,'' \emph{arXiv preprint arXiv:2305.13831}, 2023.

\bibitem{valle2020mellotron}
R.~Valle, J.~Li, R.~Prenger, and B.~Catanzaro, ``Mellotron: Multispeaker
  expressive voice synthesis by conditioning on rhythm, pitch and global style
  tokens,'' in \emph{ICASSP 2020-2020 IEEE International Conference on
  Acoustics, Speech and Signal Processing (ICASSP)}.\hskip 1em plus 0.5em minus
  0.4em\relax IEEE, 2020, pp. 6189--6193.

\bibitem{popov2021grad}
V.~Popov, I.~Vovk, V.~Gogoryan, T.~Sadekova, and M.~Kudinov, ``Grad-tts: A
  diffusion probabilistic model for text-to-speech,'' in \emph{International
  Conference on Machine Learning}.\hskip 1em plus 0.5em minus 0.4em\relax PMLR,
  2021, pp. 8599--8608.

\bibitem{lei2022cross}
Y.~Lei, S.~Yang, X.~Zhu, L.~Xie, and D.~Su, ``Cross-speaker emotion transfer
  through information perturbation in emotional speech synthesis,'' \emph{IEEE
  Signal Processing Letters}, vol.~29, pp. 1948--1952, 2022.

\bibitem{lee2021priorgrad}
S.-g. Lee, H.~Kim, C.~Shin, X.~Tan, C.~Liu, Q.~Meng, T.~Qin, W.~Chen, S.~Yoon,
  and T.-Y. Liu, ``Priorgrad: Improving conditional denoising diffusion models
  with data-dependent adaptive prior,'' in \emph{International Conference on
  Learning Representations}, 2021.

\bibitem{chou2019one}
J.~chieh Chou and H.-Y. Lee, ``{One-Shot Voice Conversion by Separating Speaker
  and Content Representations with Instance Normalization},'' in \emph{Proc.
  Interspeech 2019}, 2019, pp. 664--668.

\bibitem{Jia2018TransferLF}
Y.~Jia, Y.~Zhang, R.~J. Weiss, Q.~Wang, J.~Shen, F.~Ren, Z.~Chen, P.~Nguyen,
  R.~Pang, I.~L. Moreno \emph{et~al.}, ``Transfer learning from speaker
  verification to multispeaker text-to-speech synthesis,'' in \emph{Proceedings
  of the 32nd International Conference on Neural Information Processing
  Systems}, 2018, pp. 4485--4495.

\bibitem{lu2019one}
H.~Lu, Z.~Wu, D.~Dai, R.~Li, S.~Kang, J.~Jia, and H.~Meng, ``One-shot voice
  conversion with global speaker embeddings.'' in \emph{Interspeech}, 2019, pp.
  669--673.

\bibitem{casanova2021sc}
E.~Casanova, C.~D. Shulby, E.~G{\"o}lge, N.~M. M{\"u}ller, F.~S.~d. Oliveira,
  A.~Candido~Junior, A.~d.~S. Soares, S.~M. Alu{\'\i}sio, and M.~A. Ponti,
  ``Sc-glowtts: an efficient zero-shot multi-speaker text-to-speech model,'' in
  \emph{Proceedings}, 2021.

\bibitem{fu2019phoneme}
R.~Fu, J.~Tao, Z.~Wen, and Y.~Zheng, ``Phoneme dependent speaker embedding and
  model factorization for multi-speaker speech synthesis and adaptation,'' in
  \emph{ICASSP 2019-2019 IEEE International Conference on Acoustics, Speech and
  Signal Processing (ICASSP)}.\hskip 1em plus 0.5em minus 0.4em\relax IEEE,
  2019, pp. 6930--6934.

\bibitem{ulyanov2016instance}
D.~Ulyanov, A.~Vedaldi, and V.~Lempitsky, ``Instance normalization: The missing
  ingredient for fast stylization,'' \emph{arXiv preprint arXiv:1607.08022},
  2016.

\bibitem{shen2023naturalspeech}
K.~Shen, Z.~Ju, X.~Tan, Y.~Liu, Y.~Leng, L.~He, T.~Qin, S.~Zhao, and J.~Bian,
  ``Naturalspeech 2: Latent diffusion models are natural and zero-shot speech
  and singing synthesizers,'' \emph{arXiv preprint arXiv:2304.09116}, 2023.

\bibitem{jiang2023mega}
Z.~Jiang, J.~Liu, Y.~Ren, J.~He, C.~Zhang, Z.~Ye, P.~Wei, C.~Wang, X.~Yin,
  Z.~Ma \emph{et~al.}, ``Mega-tts 2: Zero-shot text-to-speech with arbitrary
  length speech prompts,'' \emph{arXiv preprint arXiv:2307.07218}, 2023.

\bibitem{Skerry2018Towards}
R.~Skerry-Ryan, E.~Battenberg, Y.~Xiao, Y.~Wang, D.~Stanton, J.~Shor, R.~Weiss,
  R.~Clark, and R.~A. Saurous, ``Towards end-to-end prosody transfer for
  expressive speech synthesis with tacotron,'' in \emph{international
  conference on machine learning}.\hskip 1em plus 0.5em minus 0.4em\relax PMLR,
  2018, pp. 4693--4702.

\bibitem{Wang2018Style}
Y.~Wang, D.~Stanton, Y.~Zhang, R.-S. Ryan, E.~Battenberg, J.~Shor, Y.~Xiao,
  Y.~Jia, F.~Ren, and R.~A. Saurous, ``Style tokens: Unsupervised style
  modeling, control and transfer in end-to-end speech synthesis,'' in
  \emph{International conference on machine learning}.\hskip 1em plus 0.5em
  minus 0.4em\relax PMLR, 2018, pp. 5180--5189.

\bibitem{Zhang2019LearningLR}
Y.-J. Zhang, S.~Pan, L.~He, and Z.-H. Ling, ``Learning latent representations
  for style control and transfer in end-to-end speech synthesis,'' \emph{ICASSP
  2019 - 2019 IEEE International Conference on Acoustics, Speech and Signal
  Processing (ICASSP)}, pp. 6945--6949, 2019.

\bibitem{Li2021ControllableET}
T.~Li, S.~Yang, L.~Xue, and L.~Xie, ``Controllable emotion transfer for
  end-to-end speech synthesis,'' \emph{2021 12th International Symposium on
  Chinese Spoken Language Processing (ISCSLP)}, pp. 1--5, 2021.

\bibitem{Bian2019Multi}
Y.~Bian, C.~Chen, Y.~Kang, and Z.~Pan, ``Multi-reference tacotron by intercross
  training for style disentangling,transfer and control in speech synthesis,''
  \emph{CoRR, vol. abs/1904.02373}, 2019.

\bibitem{li22h_interspeech}
T.~Li, X.~Wang, Q.~Xie, Z.~Wang, M.~Jiang, and L.~Xie, ``{Cross-speaker Emotion
  Transfer Based On Prosody Compensation for End-to-End Speech Synthesis},'' in
  \emph{Proc. Interspeech 2022}, 2022, pp. 5498--5502.

\bibitem{Li2021ControllableCE}
T.~Li, X.~Wang, Q.~Xie, Z.~Wang, and L.~Xie, ``Cross-speaker emotion
  disentangling and transfer for end-to-end speech synthesis,'' \emph{IEEE/ACM
  Transactions on Audio, Speech, and Language Processing}, vol.~30, pp.
  1448--1460, 2022.

\bibitem{diclettts}
T.~Li, C.~Hu, J.~Cong, X.~Zhu, J.~Li, Q.~Tian, Y.~Wang, and L.~Xie,
  ``Diclet-tts: Diffusion model based cross-lingual emotion transfer for
  text-to-speech — a study between english and mandarin,'' \emph{IEEE/ACM
  Transactions on Audio, Speech, and Language Processing}, pp. 1--13, 2023.

\bibitem{ganin2015unsupervised}
Y.~Ganin and V.~Lempitsky, ``Unsupervised domain adaptation by
  backpropagation,'' in \emph{International conference on machine
  learning}.\hskip 1em plus 0.5em minus 0.4em\relax PMLR, 2015, pp. 1180--1189.

\bibitem{zhu2023multi}
X.~Zhu, Y.~Lei, K.~Song, Y.~Zhang, T.~Li, and L.~Xie, ``Multi-speaker
  expressive speech synthesis via multiple factors decoupling,'' in
  \emph{ICASSP 2023-2023 IEEE International Conference on Acoustics, Speech and
  Signal Processing (ICASSP)}.\hskip 1em plus 0.5em minus 0.4em\relax IEEE,
  2023, pp. 1--5.

\bibitem{choi2021neural}
H.-S. Choi, J.~Lee, W.~Kim, J.~Lee, H.~Heo, and K.~Lee, ``Neural analysis and
  synthesis: Reconstructing speech from self-supervised representations,'' in
  \emph{Advances in Neural Information Processing Systems}, 2021, pp.
  16\,251--16\,265.

\bibitem{2018Hierarchical}
W.-N. Hsu, Y.~Zhang, R.~J. Weiss, H.~Zen, Y.~Wu, Y.~Wang, Y.~Cao, Y.~Jia,
  Z.~Chen, J.~Shen, P.~Nguyen, and R.~Pang, ``Hierarchical generative modeling
  for controllable speech synthesis,'' \emph{ArXiv}, vol. abs/1810.07217, 2019.

\bibitem{liu2021delightfultts}
Y.~Liu, Z.~Xu, G.~Wang, K.~Chen, B.~Li, X.~Tan, J.~Li, L.~He, and S.~Zhao,
  ``Delightfultts: The microsoft speech synthesis system for blizzard challenge
  2021,'' \emph{arXiv preprint arXiv:2110.12612}, 2021.

\bibitem{gulati2020conformer}
A.~Gulati, J.~Qin, C.-C. Chiu, N.~Parmar, Y.~Zhang, J.~Yu, W.~Han, S.~Wang,
  Z.~Zhang, Y.~Wu \emph{et~al.}, ``Conformer: Convolution-augmented transformer
  for speech recognition,'' \emph{Proc. Interspeech 2020}, pp. 5036--5040,
  2020.

\bibitem{zhou2021seen}
K.~Zhou, B.~Sisman, R.~Liu, and H.~Li, ``Seen and unseen emotional style
  transfer for voice conversion with a new emotional speech dataset,'' in
  \emph{ICASSP 2021-2021 IEEE International Conference on Acoustics, Speech and
  Signal Processing (ICASSP)}.\hskip 1em plus 0.5em minus 0.4em\relax IEEE,
  2021, pp. 920--924.

\bibitem{Ren2019FastSpeechFR}
Y.~Ren, Y.~Ruan, X.~Tan, T.~Qin, S.~Zhao, Z.~Zhao, and T.-Y. Liu, ``Fastspeech:
  Fast, robust and controllable text to speech,'' \emph{in Proceedings of the
  33rd International Conference on Neural Information Processing Systems},
  2019, pp. 3171–3180.

\bibitem{kong2020hifi}
J.~Kong, J.~Kim, and J.~Bae, ``Hifi-gan: Generative adversarial networks for
  efficient and high fidelity speech synthesis,'' \emph{Advances in Neural
  Information Processing Systems}, vol.~33, pp. 17\,022--17\,033, 2020.

\bibitem{zhang2023speak}
Z.~Zhang, L.~Zhou, C.~Wang, S.~Chen, Y.~Wu, S.~Liu, Z.~Chen, Y.~Liu, H.~Wang,
  J.~Li \emph{et~al.}, ``Speak foreign languages with your own voice:
  Cross-lingual neural codec language modeling,'' \emph{arXiv preprint
  arXiv:2303.03926}, 2023.

\bibitem{an2022disentangling}
X.~An, F.~K. Soong, and L.~Xie, ``Disentangling style and speaker attributes
  for tts style transfer,'' \emph{IEEE/ACM Transactions on Audio, Speech, and
  Language Processing}, vol.~30, pp. 646--658, 2022.

\end{thebibliography}
%




\end{document}